# Probability of Induced Emission in Atoms

K.V. Ivanyan*

M.V. Lomonosov Moscow State University, Moscow 119991, Russia

Expression for the probability of induced emission of high-order harmonics is obtained in the region where the multiphoton approximation is applicable to the description of the ionization of an atom. The dependence of this probability on the main parameters of the pump wave and the atomic medium is established. Criteria for observing emission are formulated with the consideration of phase locking. The possibility of amplifying a UV probe wave aimed through the region where the atoms interact with the pump wave is considered.

## I. INTRODUCTION

The generation of harmonics at frequencies corresponding to an odd number of photons of an ionizing laser wave has been investigated in several experirnental [1-4] and theoretical [5-11] studies.

In [11] and [12] the generation of high-order harmonics was studied using the analytical approach previously developed in [13] to describe the effects of the above threshold ionization of atoms. This method is based on a multiphoton mechanism for the ionization of an atom, for which the Keldysh adiabaticity parameter satisfies $\gamma > 1$. It is assumed that after the birth of a photoelectron at the threshold, it gathers additional energy as a result of repeated rescattering on the Coulomb potential of the residual parent ion, which is accompanied by the absorption of quanta of the field. In particular, the main laws governing the harmonic spectrum (the plateau and the cutoff regions) were described within this approach in [11], and the dependence of the order so of the cutoff harmonic on the intensity I of the laser wave was established. Numerical evaluations of so from the equations in [11] showed that good agreement both with the experimental results in [1] and [4] and with the theoretical theoretical calculations of other investigators [14,15] is observed in the region where the theory is applicable. For more details see [17-71].

When phase synchronism of the emitters is ensured under the conditions of an experiment, the generation of high order harmonics causes the number of photons emitted from the interaction region during a pulse of the laser wave to reach $N_\gamma \sim 10^6 - 10^7$ (this number is different for different s and depends on the intensity of the wave) [3]. The data presented in [15] imply the following dependence of $N_\gamma$ on the intensity $I$.

-----------------

*k.ivanyan@yandex.com

In the region up to photoionization saturation of the medium during a pulse ($I < I_{sat}$) the familiar power law dependence, i.e., $I_s(I) \propto I^n$ is observed for all the harmonics with small $s < n_0$ ($n_0$ is the minimal number of photons of the laser wave needed to ionize an atom). For harmonics with large $s > n_0$ this dependence is close to $I^{n_0}$ in the region indicated. Therefore, any discussion of the mechanism of harmonic generation must explain, in particular, the experimental dependence of the number of photons emitted during a pulse on the power of the ionizing wave.

Of course, under the conditions of the experiment in [3] high-order harmonics can appear as a result of both spontaneous and induced emission. Therefore, when the generation mechanism is considered, the role of the induced processes must be evaluated. This evaluation can be obtained by analyzing the possible amplification of a UV probe wave passing through the focus of the pump wave. The fundamental possibility of amplification is based on the fact that under phase-locking conditions the probabilities of harmonic generation processes are proportional to the square of the number of atoms $N$, in the interaction region, while the probability of the absorption of photons of the frequency range indicated is proportional to the first power of N, . In modern experiments on atom beams with an atomic density $n_a \approx 10^{17} - 10^{18} cm^{-3}$ this number can reach $N_a \approx 10^{10}$.

In the present work expression is obtained for the probability of the induced emission of high-order harmonics under phase-locking conditions. The dependence of these probability on the basic parameters of the pump wave and the atomic medium is established. The question of the possible amplification of a UV probe wave aimed through the interaction region is also considered.

## 2. PROBABILITY OF STIMULATED EMMISSION

The amplitude of the probability of recombination of the system to the ground state of the atoms with the emission of photons having the wave vector $\mathbf{K}$ and the frequency $\Omega$ at the time $t$ is given by the expression [12]

$$\mathbf{A}_\Omega(t) = \tilde{A}_0 e A_{0\Omega} \sum_j \exp[i(s\mathbf{k} - \mathbf{K}) \cdot \mathbf{R}_j] \xi^*(s\omega - \Omega) \exp[i(s\omega - \Omega - i\tilde{\alpha})t]. \qquad (1)$$

where $A_0$ is the amplitude of the vector potential, $\mathbf{k}$ and $\omega$ are the wave vector and the frequency of the wave, and e is the polarization unit vector (we assume that the wave is polarized linearly along the z axis: $\mathbf{e} = \mathbf{e}_z$).

The expression for the formula for the probability of a transition to a partial final state per unit time was obtained in [16] and has a form

$$\frac{d}{dt}|A_\Omega(t)| = \tilde{A}_0^{\,2}(eA_{0\Omega})^2 \left|\sum_j \exp[i(s\mathbf{k}-\mathbf{K})\cdot\mathbf{R}_j]\right|^2 2\pi\delta(s\omega-\Omega). \qquad (2)$$

Where

$$\tilde{A}_0 \frac{12}{\sqrt{2\ln 4}}\alpha(\mathbf{e}_\Omega\mathbf{e})\sqrt{\frac{\varepsilon\omega^4}{Ry^5}}J_1(z')J_{n_0/2-1}(z)\left(\sum_{n=1}^\infty \Phi(n)\sum_{q,q'=1}(-1)^q J_{2q}(z')J_{2q'}(z)\right),$$

and

$$\Phi(n) = \frac{\exp[-n/\ln(4n)]}{\sqrt{n}[1+(n\omega/Ry)^2]}\left[\frac{Ry}{\omega}\left(\frac{ez'\ln(4n)}{4n}\right)^2\right]^{n/2}$$

Let us consider the amplification effect when a weak probe wave of intensity $I_s$ with a frequency $\Omega = s\omega$ is aimed into the interaction region. We restrict ourselves to the approximation in which the field is prescribed, assuming that the increment of the number of photons over the length of the interaction region is less than the number of photons in the amplified wave. To ensure optimum amplification we assume that the probe wave is aimed strictly in the direction of propagation of the pump wave ($\theta = 0$). As is usually done in problems on induced emission, to obtain the probability of the process the delta function in (2), which gives the energy conservation law, is replaced by integration over the distributions of the interacting objects, in particular, over the spectral distribution of the intensity of the pump wave. This presumes a stationary regime for the probe wave or a high degree of monochromaticity for the external source of that wave, so that $\Delta\Omega \ll \Delta\omega \sim 1/\tau_p$, where $\tau_p$ is the pulse duration of the pump wave.

We describe the spectral distribution of the intensity of the laser wave by a simple Gaussian law with a half-width $\Delta\omega$. The probability of induced emission per unit time obtained from (2) for a system of atoms distributed over a length $x$ in the direction of propagation of the waves equals

$$w_{in}^{(s)} = \frac{(2\pi)^{3/2}}{36}\alpha\lambda_0^2\tilde{A}_0^2\frac{1}{s^3}(V_{int}n_a)^2\tau_i\left(\frac{x}{l}\right)^2\frac{\sin^2(\pi x/L_{coh})}{(\pi x/L_{coh})^2}I_0. \qquad (3)$$

We note that the probability (3) and, therefore, the increase in the intensity of the probe wave are quadratically dependent on *x.* This dependence is a consequence of the phase locking of the emitters. In the case of independent emitters the probability of the process is known to be linearly dependent on *x.*

A comparison of (12) of [16] and (3) makes it possible to relate the two probabilities:

$$w_{in}^{(s)} = w_{sp}^{(s)} \frac{\rho_0^2 \tau_i}{s^2 \omega} \left(\frac{x}{l}\right)^2 \frac{\sin^2(\pi x / L_{coh})}{(\pi x / L_{coh})^2} I_0. \tag{4}$$

We consider two limiting cases: a) $L_{coh} > l$ and b) $L_{coh} < l$. In the first case from (4) we obtain the relation between the probabilities in the entire reaction volume:

$$w_{in}^{(s)} = w_{sp}^{(s)} \frac{\rho_0^2 \tau_i}{s^2 \omega} \left(\frac{x}{l}\right)^2 I_0. \tag{5}$$

As follows for (5), the dependence of the probability of induced emission on the pump wave power in the case of phase lockin4 of all the emitters in the interaction region with the waves has the same form as for the probability of spontaneous emission ($\propto I^{n_0}$).

In the second case the factor

$$\left(\frac{x}{l}\right)^2 \frac{\sin^2(\pi x / L_{coh})}{(\pi x / L_{coh})^2}$$

in (4) is replaced by the ratio $((L_{coh}/l)^2)$, and, as a result, the induced emission increases with increasing pump wave power as $\propto I^{-n_0}$.

## 3. DISCUSSION OF THE RESULTS; CONCLUSIONS

Let us now proceed to a discussion of the results obtained for induced emission. As follows from (5), for the parameters from [3] ($\tau_p = 36 ps$ and a harmonic order s= 29) the probabilities of spontaneous and induced emission are comparable in value even when the intensity of the probe wave is $I_0 \approx 1W/cm^2$. This intensity corresponds to photons of the s-th harmonic emitted during a pulse from the interaction volume. As was noted above, the number of photons measured in [3] during a pulse for the value of s selected is $\sim 10^2$. Therefore, it should be expected that under the conditions of the experiment in [3] spontaneous and induced emission make comparable contributions to the total number of photons emitted.

To separate the induced emission and evaluate its probability, it would be reasonable to perform the following experiment. Two identical nonoverlapping atom beams with a certain distance $\delta$ between the beam centers ($\delta > 2d$) are aimed into the focus of the pump wave in a direction perpendicular to the direction of propagation of the wave. To simplify the ensuing evaluations, we shall assume that the beams are positioned symmetrically on the two sides of the center of the focus.

The number of photons of the s-th harmonic emitted during the pulse in the direction of propagation of the pump wave is measured first when one beam is injected (we denote this number by $N_{\gamma_1}$). At this point it must be proved that the quanta are emitted under phase-locking conditions. Then the number of photons of that harmonic is measured when both beams are injected simultaneously (we denote this number by $N_{\gamma_2}$). The radiation emerging from the first beam effectively plays the role of a probe wave aimed into the region where the pump wave interacts with the second beam. If the time for relaxation of the system to the ground state with the emission of photons of the harmonic under consideration, which is denoted by $\tau_r$ is less than the pulse duration $\tau_p$ of the laser wave ($\tau_r < \tau_p$), the intensity of the probe wave $I_0$, which is related to $N_{\gamma_1}$, equals

$$I_0 = \frac{N_{\gamma_1} s \omega}{\pi \rho_0^2 \tau_i}. \tag{6}$$

If the beams emit independently (if the condition $L_{coh} < \delta$ holds) the ratio between $N_{\gamma_2}$ and $N_{\gamma_1}$ can be estimated in accordance with (5) and (6) using the relation

$$\frac{N_{\gamma_2}}{N_{\gamma_1}} \approx 2 + \frac{N_{\gamma_1}}{\pi s}. \tag{7}$$

We note that this relation holds in the prescribed-field approximation for the probe wave provided that we have $\tau_r < \tau_p, \delta/c$, where the parameter $\delta/c$ defines the time needed for radiation to propagate between the beams.

In the other limiting case, $\tau_r > \tau_p, \delta/c$, the ratio $\frac{N_{\gamma_2}}{N_{\gamma_1}}$ is given by a different relation:

$$\frac{N_{\gamma_2}}{N_{\gamma_1}} \approx 2 + \frac{N_{\gamma_1}}{\pi s} \frac{\tau_i}{\tau_r}. \tag{8}$$

Finally, when the inequality $\delta/c > \tau_p, \tau_r$ holds, the beams emit independently, and
$$\frac{N_{\gamma_2}}{N_{\gamma_1}} = 2.$$

The correction to the first term in (7) and (8) associated with the induced radiation is of order unity even when we have $N_{\gamma_1} \sim 10^2$ and s=29. Hence it follows that appreciable amplification of the intensity of the high-order harmonics by the induced emission can be obtained when the experiment just described is carried out. We stress again that for (7) and (8) to be valid, the distance $\delta$ between the beams must be such that the time for propagation of a light pulse between them would be shorter than the relaxation time of the system $\tau_r$. The measurements of the number of photons obtained as a result of variation of this distance (while the remaining parameters of the problem are left unchanged) can serve as a basis for the experimental evaluation of $\tau_r$.

The relations (7) and (8) are valid when the role of the probe wave in the amplification scheme is played by spontaneous emission appearing under the condition $\delta > L_{coh} > l$. If a quasimonochromatic probe wave from an external source is used for amplification, the probability of induced emission depends on the coherence length $L'_{coh}$ of that wave. When a source with a degree of nonmonochromaticity for which $a << L'_{coh} < l < L_{coh}$ is employed, the relation (5) between the probabilities is replaced by

$$w_{in}^{(s)} = w_{sp}^{(s)} \frac{\rho_0^2 \tau_{coh}}{8s\omega} \left(\frac{L'_{coh}}{l}\right)^2 I_0. \tag{9}$$

where $\tau_{coh} \approx L'_{coh}/c$.

If the coherence length $L'_{coh}$ tends to zero (more precisely, becomes of the order of the wavelength $\lambda_0/s$ of the harmonic), the phase synchronism of the emitters in the induced processes vanishes, and the relations obtained in this work become meaningless.

**REFERENCES**

1. A. McPherson, G. Gibson, H. Jara *et al.,* J. Opt. Soc. Am. B **4,** 595 (1987).
2. M. Ferray, A. L'Huillier, L. A. Lompre *et al.,* J. Phys. **B 21,** L31 (1988).
3. L.i, A . L'Huillier, M. Ferray *et al.,* Phys. Rev. A **39,** 5751 (1989).
4. M.iy azaki and H. Sackai, J. Phys. B **25,** L83 (1992).
5. L. Pan, K. T. Taylor, and C. W. Clark, Phys. Rev. A **39,** 4894 (1989).

6. H.. Eberly, Q. Su, and J. Javanainen, Phys. Rev. Lett. **62,** 881 (1989); J. Opt. Soc. Am. B **6,** 1289 (1989).
7. K. C. Kulander and B. W. Shore, J. Opt. Soc. Am. B 7,502 (1990); Phys. Rev. Lett. **62,** 524 (1989).
8. P. B. Corkum, Phys. Rev. Lett. **71,** 1994 (1993).
9. L'.Hu illier, P. Balcou, S. Candel *er al.,* Phys. Rev. A **46,** 2778 (1992).
10. E. A. Nersesov and D. F. Zaretsky, Laser,Phys. **3,** 1105 (1993).
11. D. F. Zaretskii, E.A. Nersesov, Zh. Eksp. Teor. Fiz. **109**, 1994 (1996)
12. D. F. Zaretsky and E. A. Nersesov, Zh. Eksp. Teor. Fiz. **107,** 79 (1995).
13. D. F. Zaretsky and E. A. Nersesov, Zh. Bksp. Teor. Fiz. **103,** 1191 (1993).
14. .L . Krause, K. J. Shafer, and K. C. Kulander, Phys. Rev. Lett. 68, 3535 (1992).
15. L . Krause, K. C. Kulander, and K. J. Shafer, Phys. Rev. A **45,** 4998 (1992).
16. K.V. Ivanyan, arXiv:1709.02978 (2017).
17. K.B. Oganesyan, J. Mod. Optics, **61,** 763 (2014).
18. L.A.Gabrielyan, Y.A.Garibyan, Y.R.Nazaryan, K.B. Oganesyan, M.A.Oganesyan, M.L.Petrosyan, A.H. Gevorgyan, E.A. Ayryan, Yu.V. Rostovtsev, arXiv:1704.004730 (2017).
19. D.N. Klochkov, A.H. Gevorgyan, K.B. Oganesyan**,** N.S. Ananikian, N.Sh. Izmailian, Yu. V. Rostovtsev, G. Kurizki, arXiv:1704.006790 (2017).
20. K.B. Oganesyan, J. Contemp. Phys. (Armenian Academy of Sciences), **52,** 91 (2017).
21. Fedorov M.V., Oganesyan K.B., Prokhorov A.M., Appl. Phys. Lett., **53**, 353 (1988).
22. Oganesyan K.B., Prokhorov A.M., Fedorov M.V., Sov. Phys. JETP, **68,** 1342 (1988).
23. Oganesyan KB, Prokhorov AM, Fedorov MV, Zh. Eksp. Teor. Fiz., 53, 80 (1988).
24. A.H. Gevorgyan, M.Z. Harutyunyan, K.B. Oganesyan, E.A. Ayryan, M.S. Rafayelyan, Michal Hnatic, Yuri V. Rostovtsev, G. Kurizki, arXiv:1704.03259 (2017).
25. AS Gevorkyan, AA Gevorkyan, KB Oganesyan, Physics of Atomic Nuclei, **73**, 320 (2010).
26. D.N. Klochkov, AI Artemiev, KB Oganesyan, YV Rostovtsev, MO Scully, CK Hu, Physica Scripta, **T140,** 014049 (2010).
27. AH Gevorgyan, MZ Harutyunyan, KB Oganesyan, MS Rafayelyan, Optik-International Journal for Light and Electron, Optics, 123, 2076 (2012).
28. D.N. Klochkov, AI Artemiev, KB Oganesyan, YV Rostovtsev, CK Hu, J. of Modern Optics, **57,** 2060 (2010).
29. A.H. Gevorgyan**,** M.Z. Harutyunyan, G.K. Matinyan, K.B. Oganesyan, Yu.V. Rostovtsev, G. Kurizki and M.O. Scully**,** Laser Physics Lett., **13,** 046002 (2016).

30. K.B. Oganesyan, J. Mod. Optics, **61,** 1398 (2014).
31. AH Gevorgyan, KB Oganesyan, GA Vardanyan, GK Matinyan, Laser Physics, **24,** 115801 (2014)
32. K.B. Oganesyan, J. Contemp. Phys. (Armenian Academy of Sciences), **51,** 307 (2016).
33. AH Gevorgyan, KB Oganesyan, Laser Physics Letters **12** (12), 125805 (2015).
34. AH Gevorgyan, KB Oganesyan, EM Harutyunyan, SO Arutyunyan, Optics Communications, **283**, 3707 (2010).
35. E.A. Nersesov, K.B. Oganesyan, M.V. Fedorov, Zhurnal Tekhnicheskoi Fiziki, **56**, 2402 (1986).
36. K.B. Oganesyan, J. Mod. Optics, **62,** 933 (2015).
37. K.B. Oganesyan, M.L. Petrosyan, YerPHI-475(18) – 81, Yerevan, (1981).
38. Petrosyan M.L., Gabrielyan L.A., Nazaryan Yu.R., Tovmasyan G.Kh., Oganesyan K.B., Laser Physics, **17**, 1077 (2007).
39. A.H. Gevorgyan, K.B.Oganesyan, E.M.Harutyunyan, S.O.Harutyunyan, Modern Phys. Lett. B, **25**, 1511 (2011).
40. A.H. Gevorgyan, K.B. Oganesyan, Optics and Spectroscopy, **110**, 952 (2011).
41. K.B. Oganesyan. Laser Physics Letters, **12**, 116002 (2015).
42. GA Amatuni, AS Gevorkyan, AA Hakobyan, KB Oganesyan, et al, Laser Physics, **18,** 608 (2008).
43. K.B. Oganesyan, J. Mod. Optics, **62,** 933 (2015).
44. K.B. Oganesyan, J. Contemp. Phys. (Armenian Academy of Sciences), **50,** 312 (2015).
45. Oganesyan K.B., Prokhorov, A.M., Fedorov, M.V., ZhETF, **94**, 80 (1988).
46. E.M. Sarkisyan, KG Petrosyan, KB Oganesyan, AA Hakobyan, VA Saakyan, Laser Physics, **18,** 621 (2008).
47. DN Klochkov, KB Oganesyan, EA Ayryan, NS Izmailian, Journal of Modern Optics **63,** 653 (2016).
48. K.B. Oganesyan. Laser Physics Letters, **13**, 056001 (2016).
49. DN Klochkov, KB Oganesyan, YV Rostovtsev, G Kurizki, Laser Physics Letters **11,** 125001 (2014).
50. K.B. Oganesyan, Nucl. Instrum. Methods A **812,** 33 (2016).
51. AS Gevorkyan, AA Gevorkyan, KB Oganesyan, GO Sargsyan, Physica Scripta, **T140,** 014045 (2010).
52. AH Gevorgyan, KB Oganesyan, Journal of Contemporary Physics (Armenian Academy of Sciences) **45,** 209 (2010).
53. ZS Gevorkian, KB Oganesyan, Laser Physics Letters **13**, 116002 (2016).

54. AI Artem'ev, DN Klochkov, K Oganesyan, YV Rostovtsev, MV Fedorov, Laser Physics **17**, 1213 (2007).
55. A.I. Artemyev, M.V. Fedorov, A.S. Gevorkyan, N.Sh. Izmailyan, R.V. Karapetyan, A.A. Akopyan, K.B. Oganesyan, Yu.V. Rostovtsev, M.O. Scully, G. Kuritzki, J. Mod. Optics, **56**, 2148 (2009).
56. A.S. Gevorkyan, K.B. Oganesyan, Y.V. Rostovtsev, G. Kurizki, Laser Physics Lett., **12**, 076002 (2015).
57. Zaretsky, D.F., Nersesov, E.A., Oganesyan, K.B., and Fedorov, M.V., Sov. J. Quantum Electronics, **16**, 448 (1986).
58. K.B. Oganesyan, J. Contemp. Phys. (Armenian Academy of Sciences), **50,** 123 (2015).
59. DN Klochkov, AH Gevorgyan, NSh Izmailian, KB Oganesyan, J. Contemp. Phys., **51,** 237 (2016).
60. K.B. Oganesyan, M.L. Petrosyan, M.V. Fedorov, A.I. Artemiev, Y.V. Rostovtsev, M.O. Scully, G. Kurizki, C.-K. Hu, Physica Scripta, **T140**, 014058 (2010).
61. V.V. Arutyunyan, N. Sh. Izmailyan, K.B. Oganesyan, K.G. Petrosyan and Cin-Kun Hu, Laser Physics, **17**, 1073 (2007).
62. D.F. Zaretsky, E.A. Nersesov, K.B. Oganesyan, M.V. Fedorov, Kvantovaya Elektron. **13** 685 (1986).
63. A.H. Gevorgyan, K.B. Oganesyan, R.V. Karapetyan, M.S. Rafaelyan, Laser Physics Letters, **10**, 125802 (2013).
64. K.B. Oganesyan, Journal of Contemporary Physics (Armenian Academy of Sciences) **51,** 10 (2016).
65. M.V. Fedorov, E.A. Nersesov, K.B. Oganesyan, Sov. Phys. JTP, **31,** 1437 (1986).
66. K.B. Oganesyan, M.V. Fedorov, *Zhurnal Tekhnicheskoi Fiziki*, **57**, 2105 (1987).
67. E.A. Ayryan, M. Hnatic, K.G. Petrosyan, A.H. Gevorgyan, N.Sh. Izmailian, K.B. Oganesyan, arXiv: 1701.07637 (2017).
68. A.H. Gevorgyan, K.B. Oganesyan, E.A. Ayryan, M. Hnatic, J.Busa, E. Aliyev, A.M. Khvedelidze, Yu.V. Rostovtsev, G. Kurizki, arXiv:1703.03715 (2017).
69. A.H. Gevorgyan, K.B. Oganesyan, E.A. Ayryan, Michal Hnatic, Yuri V. Rostovtsev, arXiv:1704.01499 (2017).
70. M.V. Fedorov, K.B. Oganesyan, IEEE J. Quant. Electr, **QE-21**, 1059 (1985).
71. K.V. Ivanyan, arXiv:1708.05613 (2017).